\documentclass[%
 amsmath,amssymb,
 aps,
 pra,twocolumn
]{revtex4}

\usepackage{graphicx}
\usepackage{dcolumn}
\usepackage{bm}
\usepackage{color}

\begin{document}

\title{Reconstructing compound objects by quantum imaging with higher-order correlation functions}
\author{A.~B.~Mikhalychev$^1$, B.~Bessire$^2$, I.~L.~Karuseichyk$^1$, A.~A.~Sakovich$^1$, M.~Untern{\"a}hrer$^2$, D.~A.~Lyakhov$^3$, D.~L.~Michels$^3$, A.~Stefanov$^2$, D.~Mogilevtsev$^1$}
\affiliation{$^1$B. I. Stepanov Institute of Physics, National Academy of Sciences of Belarus, Nezavisimosti Ave. 68, Minsk 220072, Belarus;\\
$^2$Institute of Applied Physics, University of Bern, Sidlerstrasse 5, CH-3012 Bern, Switzerland;\\
$^3$Computer, Electrical and Mathematical Science and Engineering Division, 4700 King Abdullah University of Science and Technology, Thuwal 23955-6900, Kingdom of Saudi Arabia}

\date{\today}

\begin{abstract}
Quantum imaging has a potential of enhancing precision of the object reconstruction by using quantum correlations of the imaging field. This is especially important for imaging requiring low-intensity fields up to the level of few-photons.  However, quantum imaging generally leads to nonlinear estimation problems. The complexity of these problems rapidly increases with the number of parameters describing the object. We suggest a way to drastically reduce the complexity for a wide class of problems.  The key point of our approach is connecting the features of the Fisher information with the parametric locality of the problem, and building the efficient iterative inference scheme reconstructing only a subset of the whole set of parameters in each step. This iterative scheme is linear on the total number of parameters. This scheme is applied to quantum near-field imaging,   the inference procedure is developed resulting in super-resolving reconstruction of grey compound transmission objects. The functionality of the method is demonstrated with experimental data obtained by measurements of higher-order correlation functions for imaging with entangled twin-photons and pseudo-thermal light sources. By analyzing the informational content of the measurement, it becomes possible to predict  the existence of optimal photon correlations providing for the best image resolution in the super-resolution regime. This prediction is experimentally confirmed. It is also shown how an estimation bias stemming from image features may drastically improve the resolution.
\end{abstract}
\maketitle

\bigskip
\textbf{\large Introduction}

Quantum imaging implies that one uses quantum features of the imaging field source and measurement setup for enhancing precision of inferring the objects parameters from the registered image. However, for a large number of parameters describing the object, the problem of inferring these parameters from measurement results can be quite demanding, even when it is linear. It might require a prohibitive amount of measurement and computational effort to be solved. For example, to reconstruct the state of a moderate number (say, a few dozen) of the simplest quantum objects, qubits,  one already needs some simplifying assumptions, such as a low rank of the state  \cite{tao,dohoho,gross1,gross2},   the possibility to approximate the state by a matrix product  \cite{cramer}, or by a permutationally invariant state \cite{geza1,geza2,geza3}. For nonlinear problems the task is even more difficult \cite{thiebaut,nonlinear}.

Here we present an efficient method for nonlinear estimation problems that applies for the important class of parametrically localized measurements. For this class, the result of a particular measurement is dependent on a limited subset of parameters only. Such measurements are common for objects consisting of components well separated in physical or phase space. For example, measurements on individual systems in ion traps \cite{traps} or optical lattices \cite{lattice},   direct \cite{direct} and near-field imaging \cite{shih} or data-pattern tomography \cite{ourprl2010} fall into this category. For such measurements, we develop an iterative sliding window method (SWM) by reconstructing on each step only a subset of parameters which can be much smaller than the total number of parameters. The complexity for such an approach depends linearly on the number of times one needs to shift the window to cover the whole parameter set. To establish the use of the SWM, we develop an informational approach for the analysis of the measurement scheme.  We apply here the Fisher information matrix (FIM)  for the analysis of the problem and for designing the SWM. Nowadays, Fisher information analysis is firmly establishing itself as an operational tool in quantum tomography and imaging schemes \cite{tsangx,tsang16,hradil1,hradil2,seveso}. We show how the structure of the FIM can be exploited for estimating the size and structure of the parameter subset of the SWM iterations. We demonstrate the efficiency of our method with the practically important problem of imaging with correlated photons by measuring a second- or higher-order correlation function for position-momentum entangled photons generated by spontaneous parametric down-conversion (SPDC) and pseudo-thermal light. We predict the existence of an optimal degree of photon correlations in the imaging field to achieve the best resolution for a given object.  The FIM analysis allows us also to uncover the possibility to increase the resolution using biased estimation with a bias stemming from physical limitations on the set of the problem parameters.

\bigskip
\textbf{\large Results}
\medskip

\textbf{Theoretical background}
\medskip

To elucidate our approach, let us start with the simplest linear measurement model described with the probabilities $p_k=\sum\limits_{j=1}^MC_{kj}\theta_j$,
where $\theta_j$ are the parameters in question and $C_{kj}$ is the square Hermitian measurement matrix. We call the measurement strictly parametrically $l$-local, if $l<M$  and $C_{kj}=0$ for $|k-j|>l$, i.e., the matrix ${C}$ is $l$-banded, the $(l+1)$th and other side diagonals are equal to zero. It means that each probability $p_j$ depends on no more than on $2l$ neighbouring parameters. The key observation here is the possibility to approximate the inverses of banded matrices with approximately banded matrices \cite{apband,decay}. It would mean that the estimator of the parameter $\theta_j$ depends only on probabilities in the vicinity of $p_j$.  Such a locality provides the possibility of getting an accurate estimate for some $\theta_h$, for example, by minimization of the distance between a set of experimentally measured frequencies, $f_k$, $k\in [h-J,h+J]$, where $J\geq l$ is an interval around $h$, and the probabilities estimated  as $p_k\approx\sum\limits_{j=h-J}^{h+J}C_{kj}\theta_j$ \cite{cramer}.
Estimation can be performed for a sequence of $h$, thus, shifting the estimation window along the whole set of parameters. The complexity of the SWM is linear on the number of shifts required to cover the whole set of parameters. Below we elaborate on this possibility. Notice that the consideration given above holds also for non-strictly parametrically local measurements (Supplementary Note 1).

Now let us consider the general nonlinear parametric measurement model $p_k=C_k(\theta_1, \ldots\theta_M)$.
Strict parametric $l$-locality for the case would mean $\frac{\partial{p_j}}{\partial\theta_m}=0$ for $|m-j|>l$. Our suggestion is to estimate the influence of a given change of a particular parameter on the other parameters with help of the FIM and the Cramer-Rao bound (CRB). Assuming the completeness of the measurement set, $\sum\limits_{  k}p_k=1$, the FIM for this case reads
\begin{eqnarray}
F_{mn}=\sum\limits_{  j} \frac{1}{p_j}\frac{\partial{p_j}}{\partial\theta_m}\frac{\partial{p_j}}{\partial\theta_n}.
\label{fisher1}
\end{eqnarray}
For the unbiased estimate, the CRB  connects the elements of the inverse FIM
with the variance of the estimators, $\Delta^2(\theta_j)\geq [F^{-1}]_{jj}/N$, where $N$ gives the total number of events. A banded structure of the FIM would mean that an error estimate for a particular parameter, $\theta_h$, can be influenced by variations of the parameters only in some vicinity $[h-J,h+J]$. Our suggestion is to use this clue for designing the SWM as it was described above for the linear case. Also, FIM and CRB can be used for optimization of the measurement scheme aiming at lowering the error bounds per given number of measured events, $N$.
One can minimize the bound for the total measurement error described by the trace of the inverse FIM.
A banded structure of the FIM gives a clue to the connection between the width of the FIM and the total error: generally, for given diagonal elements of the FIM, increasing the width (i.e., a number and value of bands) leads to an increase of the inverse trace and the total error (Supplementary Note 2).
One can also define an empirical Rayleigh criterion for the parameter resolution from the banded structure: when the FIM is strongly diagonally dominant, $F_{jj}\gg \sum\limits_{k\neq j}|F_{jk}|$, then statistical errors for estimation of the parameter $\theta_j$ are defined  mainly by the measured $f_j$, and the parameters can be well estimated by individual measurements. Notice, that the diagonal dominance is a quite strong property imposing locality (Supplementary Note 1). For example, for a strictly $1$-banded diagonally dominant FIM, a lower bound on the variance of $\theta_j$ is defined by elements $F_{jk}$ with $|j-k|\leq2$  \cite{tridig}.

\begin{figure}[htb]
\includegraphics[width=\linewidth]{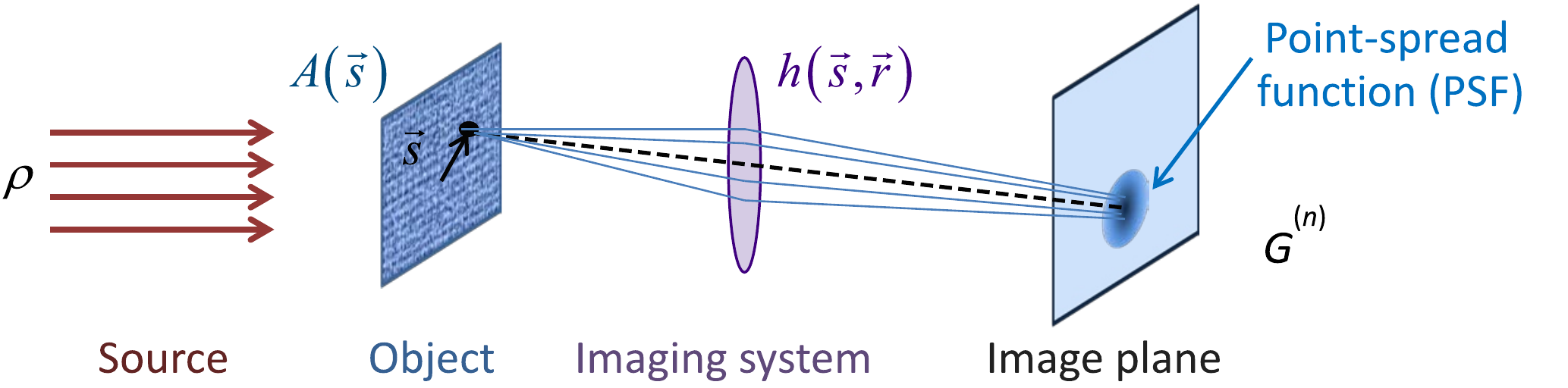}
\caption{Scheme of the measurement setup. A state of light described by the density matrix $\rho$ impinges on the object described by the transmission function $A(\vec{\mathrm{\mathbf{{s}}}})$, passes through the imaging system described by the point-spread function $h(\vec{\mathrm{\mathbf{s}}},\vec{\mathrm{\mathbf{r}}})$ and propagates to the image plane to be detected.}
\label{fig1}
\end{figure}

\textbf{Imaging with higher-order correlation functions}
\medskip

We illustrate the previous discussion applying the SWM to practically relevant examples of quantum near-field imaging by means of higher-order correlation functions. Measuring higher-order correlation functions \cite{shih,zhang,chen,zhou} is one of the ways to increase the resolution of imaging \cite{gatto,oron,oron1,classen,classen1} and to go beyond the empirical Rayleigh limit \cite{born}. The scheme of the measurement setup is schematically depicted in Fig.~\ref{fig1}. The state of linearly polarized light described by the density matrix $\rho$ impinges on the object described by the transmission function $T(\vec{\mathrm{\mathbf{s}}})$, passes through the imaging system  described by its point-spread function (PSF) $h(\vec{\mathrm{\mathbf{s}}},\vec{\mathrm{\mathbf{r}}})$ and goes to the detectors. The operator of the field amplitude, $E_{\mathrm{o}}(\vec{\mathrm{\mathbf{s}}})$, at the object plane is connected to the operator of the field amplitude at the image plane, $E(\vec{\mathrm{\mathbf{r}}})$, as
\begin{equation}
E(\vec{\mathrm{\mathbf{r}}})=\int\limits_O d^2\vec{\mathrm{\mathbf{s}}} A(\vec{\mathrm{\mathbf{s}}})h(\vec{\mathrm{\mathbf{s}}},\vec{\mathrm{\mathbf{r}}})E_{\mathrm{o}}(\vec{\mathrm{\mathbf{s}}}),
\label{eprop}
\end{equation}
where $h(\vec{\mathrm{\mathbf{s}}},\vec{\mathrm{\mathbf{r}}})$ is a PSF describing the field propagation between the object and image plane \cite{shih}. Integration in Eq.~(\ref{eprop}) is over the object plane $O$. We represent the object as a superposition of $M$ pixels
$A(\vec{\mathrm{\mathbf{s}}})=\sum\limits_{j=1}^M d_j(\vec{\mathrm{\mathbf{s}}})x_j$,
where the function $d_j(\vec{\mathrm{\mathbf{s}}})$ describes the unit transmission through the $j$th pixel, and $x_j$ is the value of the transmission assigned to the $j$th pixel. We  measure the $n$th order intensity correlation function,
$G^{(n)}_k=\mathrm{Tr}\{\left[\prod\limits_{l=1}^n{E}(\vec{\mathrm{\mathbf{r}}}_{l}^{(k)})\right]^{\dagger}\left[\prod\limits_{l=1}^n{E}(\vec{\mathrm{\mathbf{r}}}_{l}^{(k)})\right]\rho\}$,
where the index $k$ numbers some set of $n$ points, $\vec{\mathrm{\mathbf{r}}}_1^{(k)},\vec{\mathrm{\mathbf{r}}}_2^{(k)}\ldots\vec{\mathrm{\mathbf{r}}}_n^{(k)}$, in the image plane. The detection probabilities are
\begin{eqnarray}
p_k\propto \sum\limits_{  l,m} D^{(k)}(l_1\ldots l_n;m_1\ldots  m_n)
\left[\prod\limits_{i=1}^nx_{l_i}\right]^*\prod\limits_{i=1}^nx_{m_i}.
\label{pfin}
\end{eqnarray}
The coefficients $D^{(k)}(l_1\ldots l_n;m_1\ldots  m_n)$ are defined by the imaging system and the state of the source. In Supplementary Note 3, $D^{(k)}$ are derived  for SPDC entangled photons and pseudo-thermal states used for experimental implementations.

\begin{figure}[htb]
\includegraphics[width=\linewidth]{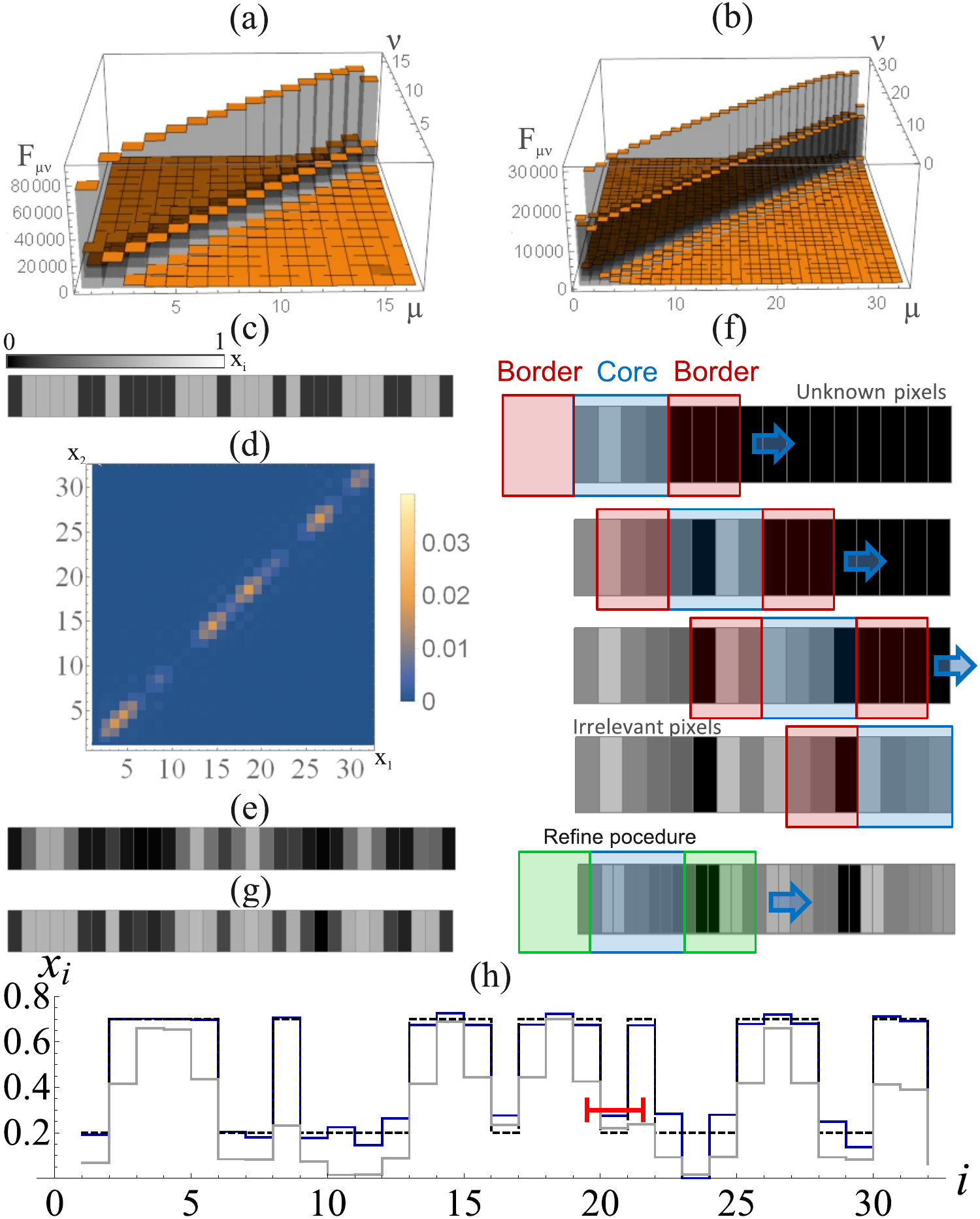}
\caption{The sliding window method. Examples of the Fisher information matrix for large pixels (a) and the Fisher information matrix for super-resolution regime (b) for reconstruction of a one-dimensional image with the second-order correlation function. Horizontal axes number the pixels. The object (c) is used for simulation of $G^{(2)}$ (panels (d) and (e) present the correlations map and its normalized diagonal part; axes in the panel (d) number the pixels), for $10^6$ joint detection events. The  sliding windows method is schematically shown in panel (f), the reconstruction result is shown in the panel (g). The result of the reconstruction (solid line) is compared to the original model object (dashed line) and diagonal part of $G^{(2)}$ (grey line) in plot (h); the horizontal axis in the panel (h) numbers pixels. Simulations are performed for a thermal source.}
\label{fig3}
\end{figure}

\bigskip
\textbf{Sliding window method for imaging}
\medskip

The problem of the object inference is to find a set of transmission values $\{x_j\}$ fitting the measured data described by the set of frequencies $f_k$ in the best way. For the realization of the SWM, we implement the following iterative scheme: On the first step, we define the  pixels (the functions $d_j(\vec{\mathrm{\mathbf{s}}})$) in such a way that the FIM, i.e.~Eq.~(\ref{fisher1}), is strongly diagonally dominant and infer the initial approximation. Then, we divide each initial pixel in a subgroup of smaller pixels, assign to each of them  the transmittance of the parent pixel and calculate the FIM. Next we define the window to be shifted as some set of adjacent ``core" pixels and some ``border" pixels around the ``core". Thereby, we use the number and relative value of major bands of the FIM for defining the size of the ``border" and perform  the fitting. Notice that for building the procedure one does not need to know the object beforehand or to perform some preliminary estimation. The pixel size and the window structure can be defined for the model object and the used imaging setup. The details of the SWM method and the pseudo-code are provided in Method section.

\medskip
\textbf{Object inference}
\medskip

Let us illustrate the mechanics of the SWM with transmitting 1D and 2D objects. We take for our examples common ``workhorses" of the quantum imaging field: a pseudo-thermal state \cite{martienssen1964} and a position-momentum entangled state produced by SPDC \cite{walborn10}. In Figs.~\ref{fig3} one can see an illustration of the SWM for a 1D object  for the simulated $G^{(2)}$ (Supplementary Note 3).  Figs. \ref{fig3}(a,b) show an example of the typical strongly diagonally dominant FIM for a pixels larger that the Rayleigh limit (Fig.~\ref{fig3}(a)) and the  FIM with pixel smaller than the Rayleigh limit (Fig.~\ref{fig3}(b)). However, the FIM of  Fig.~\ref{fig3}(b) is still narrowly banded and thus the inference problem is treatable by the SWM. The rule-of-thumb here is to choose the size of the border region larger than the number of the major bands of the FIM. Fig.~\ref{fig3}(c) shows the object.
Fig.~\ref{fig3}(d) shows the simulated $G^{(2)}$ for the object of Fig.~\ref{fig3}(c) for the thermal source. The process of reconstruction by moving the ``window" is depicted in Fig.~\ref{fig3}(f) (Methods section). The result of the reconstruction is shown in Figs.~\ref{fig3}(g,h). Notice that for the case we have come beyond the Rayleigh limit $\Delta l$, shown with the red bar in Fig.~\ref{fig3}(h): the reconstruction result \ref{fig3}(g) is close to the original object shown in Fig.~\ref{fig3}(c), while the diagonal part of the image \ref{fig3}(e) looks differently. The object inference for the higher-order correlation functions can be realized similarly to the procedure described above.

\medskip
\textbf{Experiment}
\medskip

\begin{figure}[htb]
\includegraphics[width=\linewidth]{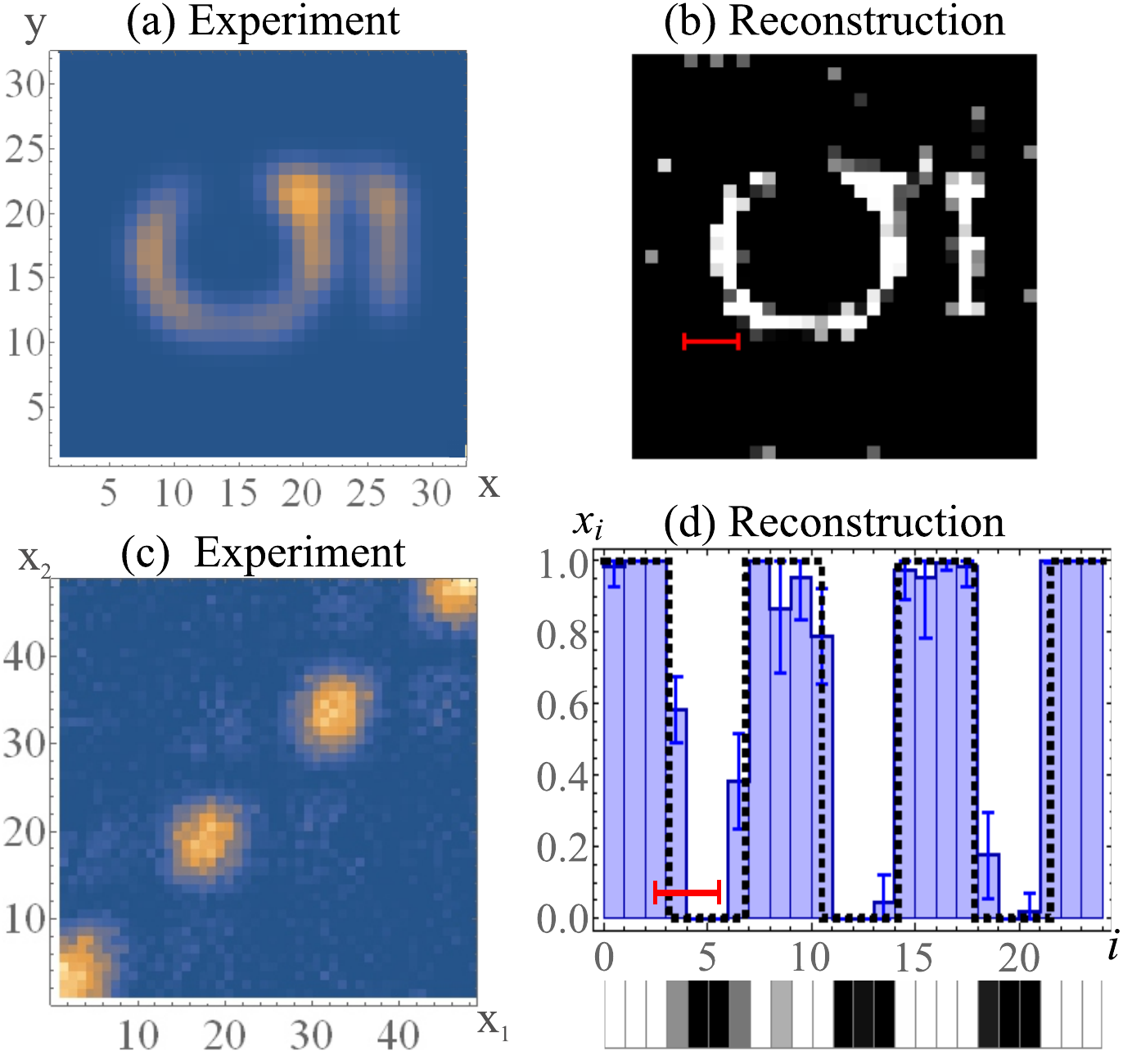}
\caption{Experimental data and reconstructed pixel transmissions. Pseudo-thermal light source and a digit ``5" object (190\,$\mu$m$\times$311\,$\mu$m) from the group 2 of a negative 1951 U.S. Air Force (USAF) resolution test chart: (a) diagonal part of a measured $G^{(3)}$ function, (b) reconstruction result. Spontaneous down-conversion source and a one-dimensional object being the positive 1951 U.S. Air Force resolution test slits  with 31.25\,$\mu$m width: (c) measured $G^{(2)}(x_1,x_2)$, (d) reconstruction result. The red segments correspond to the Rayleigh limit $\Delta l$ for the used optical system.}
\label{fig3-2}
\end{figure}

The experimental verification of the SWM was done with the particular realizations of a generic measurement scheme depicted in Fig.~\ref{fig1} for both, a pseudo-thermal and a spontaneous down-conversion (SPDC) source (Methods section). To produce pseudo-thermal light, a rotating ground glass disk was illuminated by a monochromatic laser  \cite{martienssen1964,goodman1975} operating at $405$\,nm. Type-0 position-momentum entangled two-photon states were generated by a SPDC source \cite{walborn10}. Thereby, we use a 12\,mm long periodically poled potassium titanyl phosphate (PPKTP) nonlinear crystal pumped by a continuous wave laser centered at 405\,nm. The entangled photons are then emitted at 810\,nm. Detection at the image plane was done using SuperEllen, a single photon sensitive 32$\times$32 pixel  single-photon avalanche diodes (SPAD) array detector manufactured in complementary metal–oxide–semiconductor (CMOS) technology \cite{gasparini2018,unternaehrer2018}.
Fig.~\ref{fig3-2} shows the results of the SWM for experimental data. Fig.~\ref{fig3-2}(b) shows the reconstructed 2D object inferred from the measurement of $G^{(3)}$ for the pseudo-thermal source shown in Fig.~\ref{fig3-2}(a) (only the diagonal part is shown). Figs.~\ref{fig3-2}(c, d) present reconstruction of 1D object from $G^{(2)}$ for the SPDC imaging state. Resolution beyond the Rayleigh limit (shown by red bars) is demonstrated for both sources.

\begin{figure}[htb]
\includegraphics[width=\linewidth]{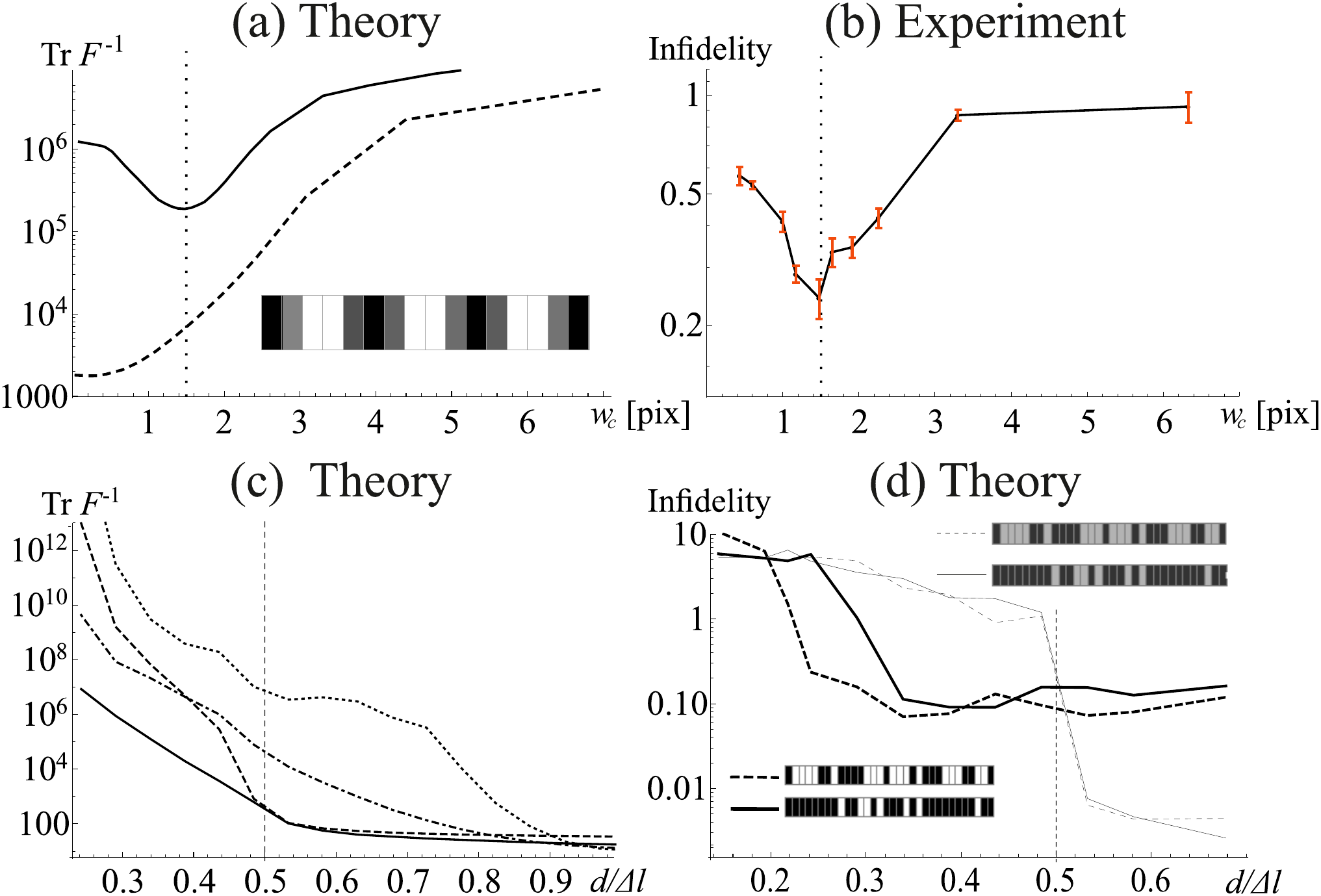}
\caption{Inverse Fisher information matrix trace  and reconstruction infidelity  for 1D pseudo-thermal light images. (a) Calculated dependence of the total measurement error on $w_c$ for the object in the inset. The solid and dashed lines correspond to $d/\Delta l=0.41,0.5$. The object pixel size $d$ is normalized by the Rayleigh limit $\Delta l$. Vertical dotted lines correspond to $w_c=1.5$ pix (the value of the minimum). (b) Measured infidelity for the same object and superresolution regime as in (a). Red bars show the standard deviations of the reconstruction results for the analysis of 12 independent 1D data sets taken from a single 2D experiment. (c) Calculated dependence of the total measurement error on $d/\Delta l$ for the top object in the insets in (d). Dotted, dot-dashed, solid, dashed lines correspond $w_c=\infty,2,1,0$ pixels. Vertical dashed lines corresponds to $d/\Delta l=0.5$. (d) Calculated infidelity for the objects in the inset. Thick lines correspond to black-and-white objects and thin lines to grey objects.}
\label{fig5}
\end{figure}
\medskip

\textbf{Optimization of the imaging state}
\medskip

The informational approach allows us to predict the optimal correlation width of the used illumination source for the object resolution in the super-resolution regime. Intuitively, it seems that the smaller the correlation width is, the better the resolution should be. However,
the analysis of the collected information shows that for the object inference perfectly correlated photons might not be the best choice. It follows from an optimization of the lower bound on the total reconstruction error. This prediction is valid for an arbitrary reconstruction method for the measurement of the second-order correlated function with both twin-photon and quasi-thermal imaging source.   In Fig.~\ref{fig5}(a) an example of the optimization is shown for the image reconstruction from a $G^{(2)}$ function for a pseudo-thermal state. The trace of the inverse FIM and the infidelity of the reconstruction are shown for different correlation widths $w_c$. There is an optimal $w_c$ allowing to increase the reconstruction quality for the same number of detector counts in the super-resolution regime. One can describe the most optimal state with the following rule-of-thumb: the correlation width should be close to the smallest object details to be resolved, i.e., to the pixel size. In the super-resolution regime, the measurement of the second-order intensity correlation gives the  most information per detected photon coincidence event about the object when the photons going through the neighbouring pixels are correlated.
This prediction is confirmed by the experimental results shown in Fig.~\ref{fig5}(b) using the pseudo-thermal source with various correlation widths of the generated speckles (Supplementary Note 3). A similar relation between the optimal photon correlation width and the size of the object features also holds for a SPDC source (Supplementary Note 4). For pixel size exceeding the Rayleigh limit this effect disappears; decreasing the correlation width brings about enhancement of the resolution (Fig. \ref{fig5}(c)).

\medskip
\textbf{Inference bias}
\medskip

The informational approach and the SWM can capture the possibility of a considerable improvement of resolution stemming from constraints imposed on the parameters. For the special case of parameters being on the borders of the allowed regions, the estimation is generally biased. The bias can significantly modify the error bounds \cite{eldarreview,eldar2004} (see Supplementary Note 5, and Fig.~\ref{fig5}(d)). For the imaging of binary black-and-white objects (i.e., for $x_j$ being either 0 or 1; see bottom inset and thick lines in Fig.~\ref{fig5}(d)), one can go far beyond the resolution limit found for grey images (top inset and thin lines  in Fig.~\ref{fig5}(d)) even without any prior assumption of the binary object structure. The reason for it is the dependence of the errors bounds on the bias derivative with respect to the parameters \cite{eldarreview}. Generally, the SWM shifts the estimators near borders. The closer the estimated value is to the border, the larger is the respective shift and the error bound deviation. Notice that for the object inference demonstrated in Fig.~\ref{fig3-2}, this bias effect was actually seen.

\bigskip
\textbf{\large Discussion}
\bigskip

We developed an inference  method for nonlinear parametrically local problems and showed how the analysis of the information allows one to develop an estimation scheme making the complexity of the problem linear on the total number of parameters. Then,  the scheme was applied to the experimental data for superresolution imaging based on higher-order correlation measurements with non-classical two-photon and pseudo-thermal states. It was shown how the FIM can be applied to optimize the imaging state for better resolution, in particular, for the correlation width of the twin-photon or pseudo-thermal imaging fields. Generally, the correlation width should be close to the smallest details to be resolved. This prediction is experimentally confirmed for measurements with pseudo-thermal light. It was also demonstrated that bias due to marginal values of estimated parameters can improve the resolution. We believe that the suggested SWM and an information approach for nonlinear inference problems will find applications for the design and optimization of inference schemes in imaging, quantum diagnostics and tomography.
\bigskip

\textbf{\large Methods}
\medskip

\textbf{The sliding windows method}
\medskip

The practical application of the SWM to the quantum imaging problem consists of the following steps.

First, an initial rough estimate of the object transmission amplitude is found. The pixel size $d_{\mathrm{initial}}$ is chosen in such a way that the inverse of the FIM for the reconstruction of the object, expressed in terms of pixels of the size $d_0^{(0)}$, is diagonally dominant. The problem is strongly local and a single run of the SWM is sufficient for getting the initial estimate.

Then, the estimate is refined by representing the object in terms of smaller pixels (size $d$) and applying the reconstruction algorithm again. The pixel size $d$ limits the size of the object features that can be successfully reconstructed and, therefore, determines the achievable resolution.

\begin{figure}
  \centering
  \includegraphics[width=\columnwidth]{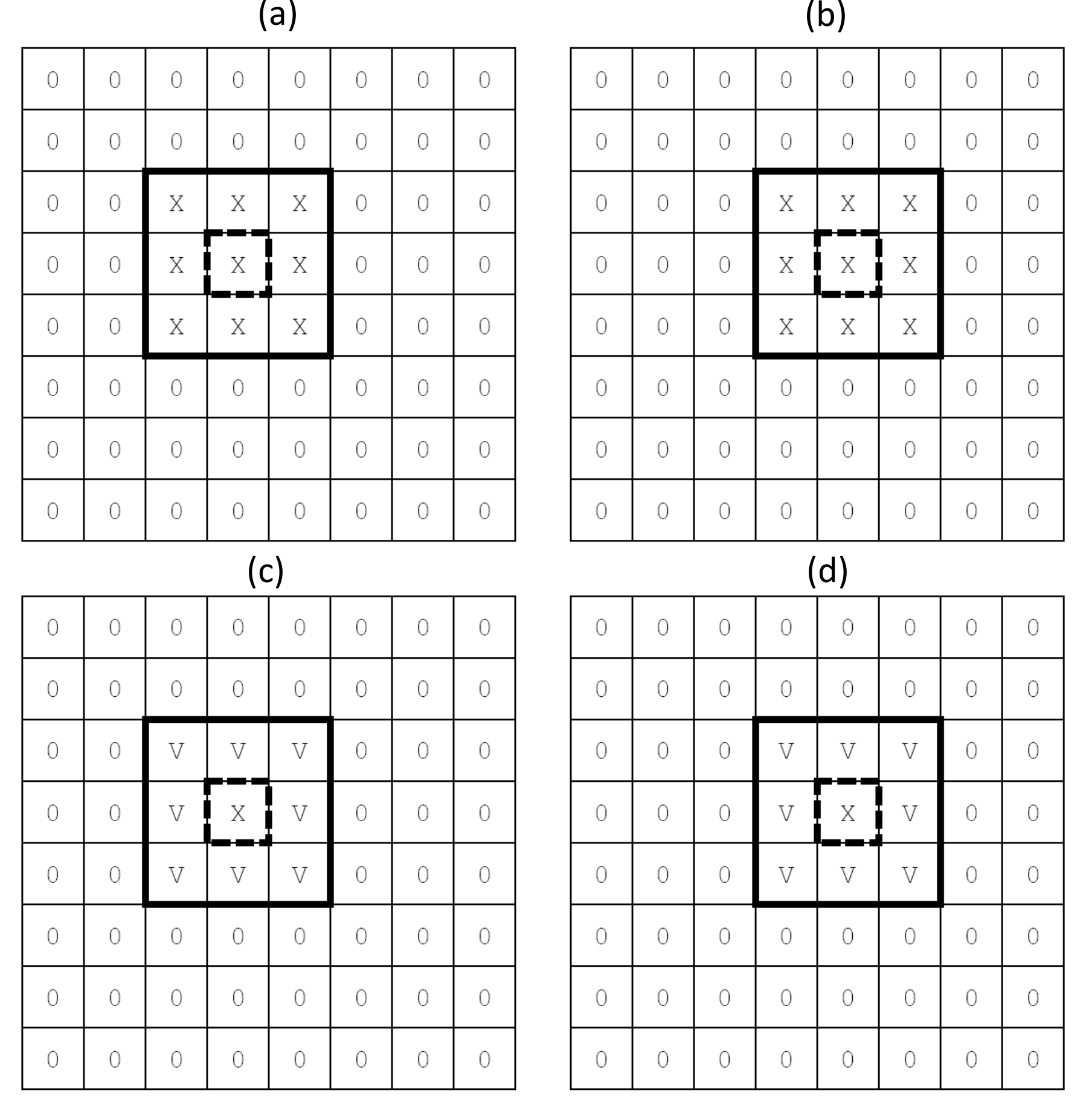}
  \caption{An example of two adjacent reconstruction steps for a general two-dimensional case. Figs.~(a,b) correspond to obtaining the first approximation and Figs.~(c,d)  correspond to iterative refinement. Although pixels are divided after the first approximation is obtained, here the pixel size is chosen to be the same for both cases in order to show mathematical similarity of the first approximation inference and refinement. Letter ``\texttt X'' denotes pixels for which the optimization problem will be stated (``unknown'' pixels), digit ``\texttt 0'' denotes pixels which will be completely ignored at the iteration (``irrelevant'' pixels), and letter ``\texttt V'' denotes pixels whose values will be included in the optimization problem as known constants (``known'' pixels). Dashed frame shows the core of the window, solid frame shows the whole window, the core and the border.}\label{Appendix.Pictures.WinRecAdjacentStepsBoth}
\end{figure}

Here the pseudo code for the iterative reconstruction is presented. Both algorithms (the one for the first approximation inference and the one for refinement) are mathematically very similar, the only difference is the role of the window border: the first algorithm implies that pixels inside the border are unknown and not reliable (at each step): they need to be reconstructed but then discarded; the second algorithm implies that pixels inside the border are known (at each step) and thus can be included in the optimization problem as known constants (Fig. \ref{Appendix.Pictures.WinRecAdjacentStepsBoth}).

The pseudo code for the first approximation inference algorithm reads as follows:
\bigskip

  For given core window dimensions, compute all core window positions which lead to the object being fully covered by non overlapping core windows.
 \bigskip

 For each position of the core window:
 \bigskip

   For a given border size, build a complete window: core + border.
   \bigskip

    Map the resulting full window to the image plane.
    \bigskip

     Find all detectors inside the obtained window in the image plane.
    \bigskip

    Combine the obtained detectors according to the order of correlations and other constraints if present.
    \bigskip

     For the obtained detector combinations, load corresponding experimental joint detection frequencies.
   \bigskip

    Build a function that maps pixels inside the full window to residual between theoretical joint detection probabilities and experimental frequencies. The theoretical probabilities are computed assuming all pixels but those inside the window to be zero.
    \bigskip

    Perform numerical minimization of the function thus obtained. Pixels inside the full window are subjected to physical constraints.
   \bigskip

    Update object pixels inside the core window with the corresponding values obtained from the minimization procedure. Discard other pixel values.

\bigskip

Because the algorithm first computes all core window positions and then applies one iteration per core window position, it is easily paralleled.

The pseudo code for the iterative refinement algorithm reads as follows:
\bigskip

For a given core window position and a given border size, build a complete window: core + border.
\bigskip

   Map the resulting window to the image plane.
  \bigskip

  Find all detectors inside the obtained window in the image plane.
  \bigskip

   Combine the obtained detectors according to the order of correlations and other constraints if present.
   \bigskip

  For the obtained detector combinations, load corresponding experimental joint detection frequencies.
  \bigskip

 Build a function that maps pixels inside the core window to the residuals between theoretical joint detection probabilities and experimental frequencies. The theoretical probabilities are computed assuming pixels outside the core window but inside the full window to be known, they are set to constant values. Pixels outside the full window are set to zero.
 \bigskip

  Perform numerical minimization of the function thus obtained. Pixels inside the core window are subject to physical constraints.
  \bigskip

   Update pixels inside the core window with the values obtained.
   \bigskip

 Move the core window one step further.

\bigskip

\begin{figure}[htbp]
\centering\includegraphics[width=8.5cm]{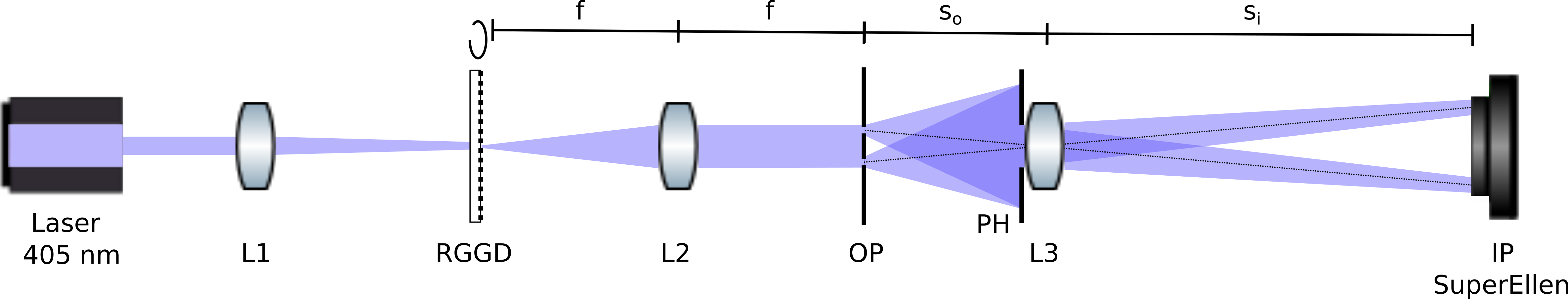}
\caption{\label{fig:thermal_setup}Pseudo-thermal light imaging setup. A monochromatic laser is focused onto a rotating ground glass disk (RGGD) by means of lens L1. The subsequent lens L2 provides the far-field speckle pattern at the object plane (OP). The resolution  of the single lens (L3) imaging system can be modified by a variable size pinhole (PH). Single photons are detected at the image plane (IP) by SuperEllen.}
\end{figure}
\begin{figure}[htbp]
\centering\includegraphics[width=8.5cm]{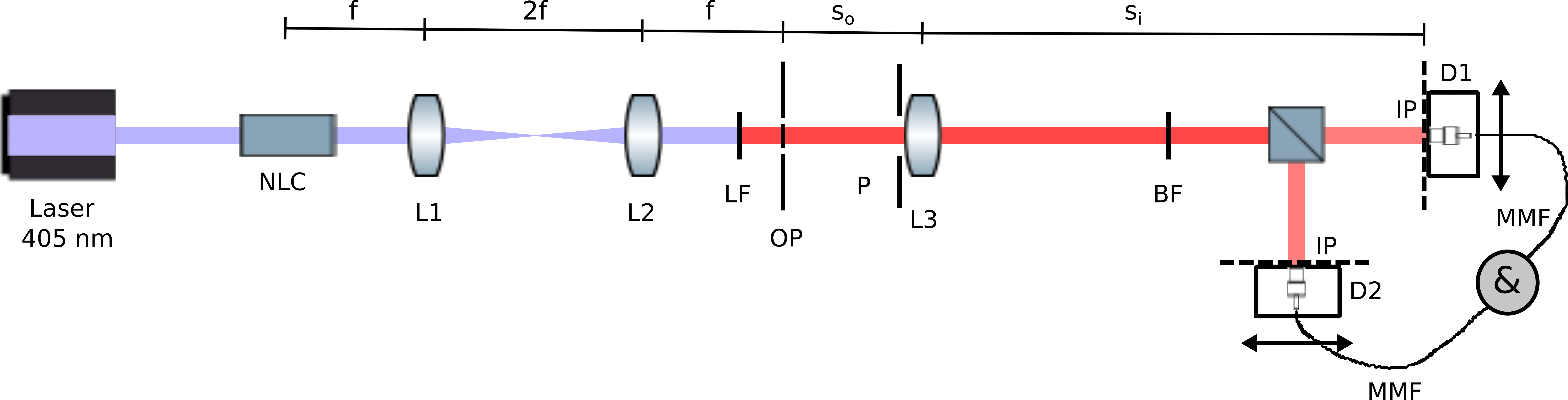}
\caption{SPDC setup. A monochromatic laser is weakly focused into a PPKTP nonlinear crystal (NLC) to generate type-0 SPDC. The two-photon state is imaged via a 4-$f$ arrangement from the center of the NLC to the object plane (OP) using lenses L1 and L2. A long-pass filter (LF) blocks the pump and a band-pass filter (BF) transmits photons at 810\,nm. A single lens imaging system with lens L2 maps the OP onto the image plane (IP) which coincides with the fiber tip of two multimode fibers (MMFs) connecting the detection stages D1 and D2 in a coincidence circuit. The resolution of the imaging system is modified by a variable size pinhole (PH).  \label{fig:setup_SPDC}}
\end{figure}

\medskip
\textbf{Experiments}
\medskip

In the first experiment we illuminate a rotating ground glass disk (RGGD) with an attenuated, monochromatic laser operating at $\lambda=405$\,nm (Fig.~\ref{fig:thermal_setup}). An additional lens (L1) in front of the disk allows to vary the beam waist radius at the position of the RGGD. Subsequently, we insert a far-field lens (L2) in a $2f$ setting ($f=75$\,mm) in order to collimate the light and remove the spherical wave-front given by the point-like source. (The latter has shown to induce distortions in the subsequent imaging setup.) The object plane (OP) is then located in the far-field  of the source.
An object is then imaged onto the image plane (IP) by means of L3 ($f=150$\,mm) which is additionally endowed with a variable size pinhole (PH) to control the resolution, i.e.~the Rayleigh limit of the setup. The diameter of the PH was fixed to 1.7\,mm. The magnification factor $m=s_i/s_o$ of the imaging system is $m=1.94$ whereas the object distance is $s_o=234$\,mm and the imaging distance is given by $s_i=454$\,mm.
At the image plane, photons are detected by SuperEllen, a single photon sensitive 32$\times$32 pixel SPAD array detector manufactured in CMOS technology with a pixel pitch of 44.64\,$\mu$m and a fill-factor of 19.7\% \cite{gasparini2018,unternaehrer2018}. SuperEllen is able to provide frames with a data acquisition window of 30\,ns and a readout time of  10\,$\mu$s at a frame rate of 800\,kHz. The spatial correlations between pixels were evaluated between consecutive frames with a resolution of 10\,$\mu$s given by the frame separation. This procedure allows for the resolution of the coherence time of the speckles of the order of $\mu$s. Second- and third-order correlation functions are measured with SuperEllen.

The here presented pseudo-thermal light setup was used to obtain the following two results: Firstly, the digit ``5" (Group 2) from a negative U.S. Air Force (USAF) test chart was imaged and then reconstructed from the data of a $G^{(3)}$ function measurement. This is shown in Fig.~3
~(a) and (b) of the main text. Secondly,  a  negative USAF chart 3-slit pattern (Group 3, Element 2) was imaged from the object plane to the image plane for various correlation widths $w_c$. The latter was modified by changing the distance between the focusing lens L1 and the RGGD and therefore the beam waist radius. Based on a $G^{(2)}$ measurement this allowed to demonstrate the dependence of the image reconstruction quality on the correlation width of the source shown in Fig.~4
~(b) of the main text.

The setup for imaging with entangled photons is shown in Fig.~\ref{fig:setup_SPDC}.  Our source generates type-0 position-momentum entangled photon states by pumping a 12\,mm long PPKTP nonlinear crystal (NLC) with a continuous wave (CW) laser centered at 405\,nm \cite{walborn10}. The entangled photons are then emitted at 810\,nm. The residual pump beam is subsequently blocked by a long-pass filter (LF) and the subsequent band-pass filter (BF) transmits photons at 810\,nm with a spectral full width at half maximum (FWHM) of 10\,nm to the detectors. The experimental setup contains two imaging systems: The first system consists of a 4-$f$ image using lenses $L1$ and $L2$ both with focal length $f=50$\,mm. This configuration maps the entangled photon states transverse momentum distribution from the OP1 at the center of the NLC to the OP with a magnification factor of $m=1$. The OP is then imaged with a single lens system onto the fiber tips of two multimode fibers (MMFs). Thereby, we have a magnification factor of $m=12$ for $s_o=65$\,mm and $s_i=780$\,mm. Both detection stages can be scanned in horizontal direction.  This setup was used to record the image of a three-slit pattern of a positive USAF resolution chart (Group 4, Element 1) by measuring a second-order correlation function of the photons. The experimental correlation map and the reconstructed object can be seen in Fig.~3
(c),(d) of the main text.

\bigskip

\textbf{Data availability}
\medskip

The datasets generated and analysed during the current study are available from the depository of the Center for Quantum Optics and Quantum Information B. I. Stepanov Institute of Physics, National Academy of Sciences of Belarus http://master.basnet.by/Informational-approach-for.data.rar .  The code itself  is available upon request.

\section*{ Supplementary note 1.  Bounds for inverses of banded and approximately banded matrices}

Here we give a number of known results about inverses of banded and approximately banded matrices useful for our discussion.
First of all, inverses of banded matrices can be approximated by banded matrices. For the inverse, $A^{-1}$, of the $l$-banded matrix $A$, one can always find such a $n*l$-banded matrix $B$ such that
\begin{equation}
\mathrm{dist}(A^{-1},B)\leq \frac{1}{|\lambda_{min}|}\left(\frac{|\lambda_{max}|-|\lambda_{min}|}{|\lambda_{max}|+|\lambda_{min}|}\right)^{n+1},
\label{bound1}
\end{equation}
where the distance is defined as $\inf\limits_{B\in B^{n*l}}\|A-B\|$, where $B^{n*l}$ is the set of all $n*l$-banded matrices, and the norm is defined as the maximal eigenvalue, $\|A\|\equiv|\lambda_{max}|$; $\lambda_{min}$ is the minimal eigenvalue of the matrix $A$  \cite{apband}.

Now we introduce the concept of the  approximately banded matrix \cite{apband}.  For a  given invertible matrix $A$ with $C=A^{-1}$ we introduce sets of all $l$-banded matrices $A_l$,  and the distances $\mathrm{dist}(A,A_l)=\delta_l$.  We call the matrix $A$  approximately banded if distances $\delta_l$ tend to zero with increasing $l$. For infinite matrices this condition can be formalized as
\begin{equation}
\lim\limits_{l\rightarrow \infty}\mathrm{dist}(A,A_l)=0.
\end{equation}
For infinite matrices the class of the approximately banded matrices is closed with respect to inversion. Notice that if the bandwidth of both, the direct and inverse matrices is much smaller than the matrix size, then the conclusions derived for infinite matrices will obviously hold for the finite ones. For approximately banded matrices, the bound
\begin{eqnarray}
\mathrm{dist}(C,A_{3lk})\leq \frac{2\delta_l\alpha_k}{|\lambda_{min}|^2}+\frac{(\kappa_l^+)^2}{|\lambda_{max}|+2\delta_k}\left(\frac{(\kappa_l^+)^2-1}{(\kappa_l^-)^2+1}\right),
\label{aband}
\end{eqnarray}
where $\alpha_k=|\lambda_{min}|/(|\lambda_{min}|-2\delta_k)$, and
\[\kappa_k^{\pm}=\frac{|\lambda_{max}|\pm2\delta_k}{|\lambda_{min}|-2\delta_k}
\]
holds \cite{apband}. A number of strong and illustrative results exists for inversions of strictly diagonally dominant banded matrices. The matrix $A$ is strictly diagonally dominant if $|a_{jj}|>\sum\limits_{j\neq k} |a_{jk}|$. For a 1-banded (tridiagonal) real square matrix $A$  there is the bound
\begin{eqnarray}
\nonumber
\left(|a_{j,j}|+s_j|a_{j,j-1}|+t_j|a_{j,j+1}|\right)^{-1}\leq |c_{j,j}|\leq
\\
\left(|a_{j,j}|+f_j|a_{j,j-1}|+g_j|a_{j,j+1}|\right)^{-1}
\label{tri1}
\end{eqnarray}
for the diagonal elements of the inverse matrix \cite{liu,tridig}. For simplicity sake, here we give the coefficients $s_j$, $t_j$, $f_j$, $g_j$ for the matrix $A$ with non-negative coefficients
\begin{eqnarray}
\nonumber
s_j=-\frac{|a_{j-1,j}|}{|a_{j-1,j-1}|+|a_{j-1,j-2}|}, \\
 f_j=-\frac{|a_{j-1,j}|}{|a_{j-1,j-1}|-|a_{j-1,j-2}|}\\
\nonumber
t_j=-\frac{|a_{j+1,j}|}{|a_{j+1,j+1}|+|a_{j+1,j+2}|}, \\
\nonumber
g_j=-\frac{|a_{j+1,j}|}{|a_{j+1,j+1}|-|a_{j+1,j+2}|}.
\label{tri2}
\end{eqnarray}
The most remarkable observation about Eqs.~(\ref{tri1},\ref{tri2}) relevant for the context of this work is that the $j$-th diagonal elements of the inverse matrix are bounded by
elements of the original matrix in the vicinity $j\pm2$.  This holds for arbitrary real invertible matrix $A$ as well. Notice, that the bound Eq.~(\ref{tri1}) does not imply that the inverse matrix is close to the tridiagonal one. The similar bound exists for pentadiagonal matrices \cite{pentadig}. In this case the $j$th diagonal elements of the inverse matrix are bounded by elements of the original matrix in the vicinity $j\pm3$.

\section*{Supplementary note 2. The Fisher information matrix and the scheme optimization}

Here, we show how the error estimation given by the Cramer-Rao bound is affected by the banded structure of the Fisher information matrix (\ref{fisher1}).
In a standard way, the overall performance of the image reconstruction scheme can be characterized with the lower bound on the total variance of the estimated parameters
\begin{equation}
\Delta^2_{tot}=\sum\limits_{j=1}^M\Delta^2_j\geq \mathrm{Tr}\left[{ F}^{-1}\right]\frac{1}{N}.
\label{total}
\end{equation}

The trace of the inverse matrix is majorized by the  smallest eigenvalue of the Fisher matrix,
\[1/{\lambda}_{min}\leq \mathrm{Tr}\left[{F}^{-1}\right]\leq M/{\lambda}_{min},\] where
${\lambda}_{min}$ is the eigenvalues of the matrix ${F}$ in Eq.~(\ref{fisher1}), and $M$ is the number of parameters.

The fact that the number and value of off-diagonal elements affects the eigenvalues and the total error,  can be well attested by the "Gershgorin's circles" theorem. It states that
\begin{equation}
\lambda_{min}\geq \min\limits_{  j}\left\{{F}_{jj}-\sum\limits_{k\neq j}|{F}_{jk}|\right\},
\label{sum}
\end{equation}
i.e., the lower bound on the lowest eigenvalue decreases with increasing off-diagonal elements.

There is another  simple and illustrative bound on the lower eigenvalue of the Fisher matrix demonstrating how the deviation of this matrix from being the diagonal one affects the error. For a general complex $M\times M$ matrix ${F}$ with real eigenvalues, the bound
\begin{eqnarray}
\nonumber
\lambda_{min}\geq \frac{\mathrm{Tr}\{F\}}{M}-\sqrt{(M-1)S}, \\
S=\frac{\mathrm{Tr}\{F^2\}}{M}-\left[\frac{\mathrm{Tr}\{F\}}{M}\right]^2
\label{tracebound}
\end{eqnarray}
holds \cite{traces}. For the trace of the squared matrix one has ${\mathrm{Tr}\{F^2\}}=\sum\limits_{j,k=1}^M F_{jk}F_{kj}$. So, for the constant trace, larger off-diagonal elements lead to lowering the bound in Eq.~(\ref{tracebound}).

For application in near-field imaging we introduce the Fisher information matrix in the following way taking into account the physical situation. We assume that $P_S$ is the sum of probabilities for the object being completely transparent in the whole object plane, taking that $P_S>\sum\limits_{  k}{ p}_k$ for all images that we are considering. Then, we take all the probabilities ${\bar p}_k=p_k/P_S$ and add the "no counts" probability $P_0=1-\sum\limits_{  k}{\bar p}_k$ to form a complete set. For simplicity sake we assume the transmissions to be real, and write down the Fisher information matrix
\begin{eqnarray}
\nonumber
{\bar F}_{ml}={F}_{ml}+
\frac{1}{P_0}\frac{\partial P_0}{\partial x_m}\frac{\partial P_0}{\partial x_l}, \\
{F}_{ml}=\sum\limits_{  k}\frac{1}{{\bar p}_k}\frac{\partial {\bar p}_k}{\partial x_m}\frac{\partial {\bar p}_k}{\partial x_l}
\label{fisher2}
\end{eqnarray}
as the sum of the matrix ${F}_{ml}$ and the rank one matrix. Here, we can use $F$ instead of $\bar F$ in the bound (\ref{total}), since
for the non-negative matrices an addition of the rank one positive matrix increases the smallest eigenvalue.

\section*{Supplementary note 3. Estimation of $D^{(k)}$ coefficients for near-field imaging}

Here we elaborate on the coefficients 
connecting the registered values of the correlation function, the density matrix of the imaging state, and the characteristics of the imaging setup. The detection probabilities can be represented as polynomials on the transmissions
\begin{eqnarray}
p_k\propto \sum\limits_{  l,m} D^{(k)}(l_1\ldots l_n;m_1\ldots  m_n)
\left[\prod\limits_{i=1}^nx_{l_i}\right]^*\prod\limits_{i=1}^nx_{m_i},
\label{pfin_appendix}
\end{eqnarray}
where the parameters are given by
\begin{equation}
\begin{gathered}
D^{(k)}(l_1\ldots l_n;m_1\ldots  m_n) \\ {} =
\mathrm{Tr}\left\{\left[\prod\limits_{i=1}^n E_{l_i}({\vec r}_i^{(k)})\right]^{\dagger}\left[\prod\limits_{j=1}^n E_{m_j}({\vec r}_j^{(k)})\right]\rho\right\}
\label{dgeneral}
\end{gathered}
\end{equation}
and
\[E_{l_i}({\vec r}_i^{(k)})=\int\limits_{OP}d^2 \vec s\, d_{l_i}({\vec s})E_o({\vec s}) h(\vec s, \vec r_i^{(k)}).
\]
The parameters $D^{(k)}(l_1\ldots l_n;m_1\ldots  m_n)$ completely describe our measurement setup. Further, they are independent of the image and they have a number of general properties stemming from the features of the point-spread function (PSF), $h(\vec s, \vec r_i^{(k)})$, and the features of the field described by the density matrix $\rho$.

Here, we give the parameters $D^{(k)}(l_1\ldots l_n;m_1\ldots  m_n)$ for the two imaging states, used experimentally, and several correlation functions.

\subsection*{Pseudo-thermal states}

It is well-known that a rotating ground glass disk (RGGD) illuminated by a laser beam provides a pseudo-thermal light source via a randomized speckle pattern \cite{martienssen1964,goodman1975}. Thereby, in the far-field of the source, the spatial correlation width $w_{c}$, i.e.~the average size of a speckle, is inversely proportional to the spatial extent of the source
(the beam waist radius on the RGGD).
Inside a speckle, the light is fully coherent whereas two different speckles are mutually uncorrelated. On the other hand, the coherence time of the field, i.e.~the speckle lifetime, is, in addition to beam waist radius, related to the rotation velocity of the RGGD: The faster the disk rotates, the shorter becomes the coherence time of the speckle. In the following we consider coincidence measurements within the coherence time of a speckle.
To estimate the actual correlation width $w_c$, we measure a second-order correlation function $G^{(2)}(\vec r_{1},\vec r_{2})$ and fit the latter by a Gaussian shape
\begin{equation}\label{eq:lcff}
G^{(2)}(\vec r_{1},\vec r_{2}) \propto \exp \left(- 2 \frac{\left| \vec r_{1}-\vec r_{2}\right|^2}{(m w_c)^2} \right),
\end{equation}
where $m$ is the magnification factor of the optical system used.

In the here presented first experiment we illuminate a RGGD with an attenuated, monochromatic laser operating at $\lambda=405$\,nm (Fig.~\ref{fig:thermal_setup}). An additional lens (L1) in front of the disk allows to vary the beam waist radius at the position of the RGGD. Subsequently, we insert a far-field lens (L2) in a $2f$ setting ($f=75$\,mm) in order to collimate the light and remove the spherical wave-front given by the point-like source. (The latter has shown to induce distortions in the subsequent imaging setup.) The object plane (OP) is then located in the far-field  of the source.
An object is then imaged onto the image plane (IP) by means of L3 ($f=150$\,mm) which is additionally endowed with a variable size pinhole (PH) to control the resolution, i.e.~the Rayleigh limit of the setup. The diameter of the PH was fixed to 1.7\,mm. The magnification factor $m=s_i/s_o$ of the imaging system is $m=1.94$ whereas the object distance is $s_o=234$\,mm and the imaging distance is given by $s_i=454$\,mm.

\begin{figure}[htbp]
\centering\includegraphics[width=8.5cm]{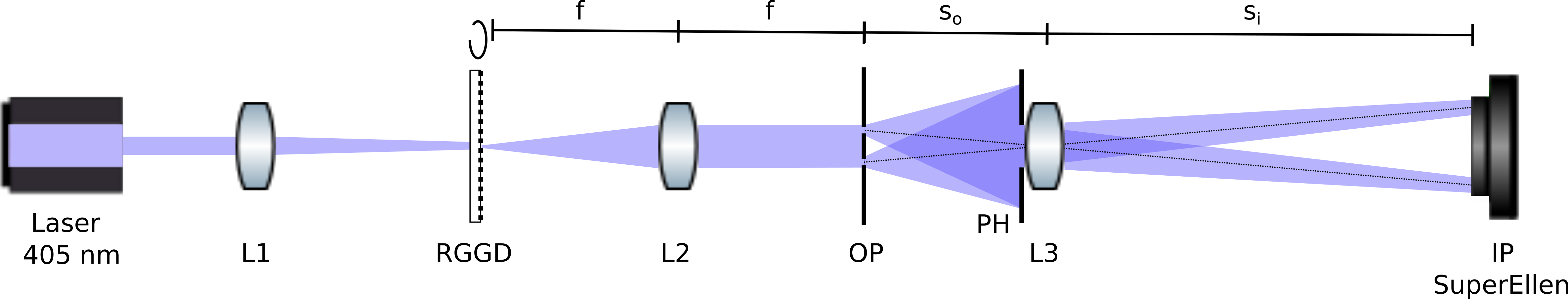}
\caption{\label{fig:thermal_setup}Pseudo-thermal light imaging setup. A monochromatic laser is focused onto a rotating ground glass disk (RGGD) by means of lens L1. The subsequent lens L2 provides the far-field speckle pattern at the object plane (OP). The resolution  of the single lens (L3) imaging system can be modified by a variable size pinhole (PH). Single photons are detected at the image plane (IP) by SuperEllen.}
\end{figure}

At the image plane, photons are detected by SuperEllen, a single photon sensitive 32$\times$32 pixel SPAD array detector manufactured in CMOS technology with a pixel pitch of 44.64\,$\mu$m and a fill-factor of 19.7\% \cite{gasparini2018,unternaehrer2018}. SuperEllen is able to provide frames with a data acquisition window of 30\,ns and a readout time of  10\,$\mu$s at a frame rate of 800\,kHz. The spatial correlations between pixels were evaluated between consecutive frames with a resolution of 10\,$\mu$s given by the frame separation. This procedure allows for the resolution of the coherence time of the speckles of the order of $\mu$s. Second- and third-order correlation functions are measured with SuperEllen.

The here presented pseudo-thermal light setup was used to obtain the following two results: Firstly, the digit "5" (Group 2) from a negative USAF chart was imaged and then reconstructed from the data of a $G^{(3)}$ function measurement. This is shown in Fig.~3
~(a) and (b) of the main text. Secondly,  a  negative USAF chart 3-slit pattern (Group 3, Element 2) was imaged from the object plane to the image plane for various correlation widths $w_c$. The latter was modified by changing the distance between the focusing lens L1 and the RGGD and therefore the beam waist radius. Based on a $G^{(2)}$ measurement this allowed to demonstrate the dependence of the image reconstruction quality on the correlation width of the source shown in Fig.~4
~(b) of the main text.

Assuming that the statistics of the pseudo-thermal source is Gaussian one can express the $n$th order correlation function, measured at points $\vec r_{j_1}$, \ldots, $\vec r_{j_n}$, as a combination of pairwise correlations
\begin{equation}
\label{eqn:Gn_thermal}
G^{(n)}(\vec r_{j_1},\ldots,\vec r_{j_n}) \propto  \sum_{(i_1,\ldots, i_n) \in S(j_1,\ldots,j_n)} \prod_m I(\vec r_{j_m}, \vec r_{i_m}),
\end{equation}
where $S(j_1,\ldots,j_n)$ represents the set of all permutations $(i_1,\ldots,i_n)$ of numbers $(j_1,\ldots,j_n)$ and
\begin{equation}
\label{eqn:I_general}
\begin{gathered}
I(\vec r, \vec r') = \int d^2 \vec s d^2 \vec s' \exp\left(-|\vec s - \vec s'|^2 / w_c^2\right) \, A(\vec s') \\ {} \times [A(\vec s)]^\ast h^\ast(\vec s, \vec r) h(\vec s', \vec r').
\end{gathered}
\end{equation}
The PSF for the here discussed near-field imaging scheme is a product $h(\vec s, \vec r) = h_0(\vec s, \vec r) \Xi(\vec s, \vec r)$ of the jinc function, also known as sombrero function \cite{shih},
\[h_0(\vec s, \vec r)=2J_1(x)/x, \quad x=\frac{R\omega}{s_oc}\left|\vec{s}+\vec{r}\frac{s_o}{s_i}\right|,
\]
and a phase factor
\[\Xi(\vec s, \vec r)=\exp\Bigl\{\frac{i\omega}{2cs_o}|\vec s|^2 + \frac{i\omega}{2cs_i}|\vec r|^2)\Bigr\},
\]
where $J_1(x)$ is the first-order Bessel function, $\omega$ is the frequency of the imaging field  and $R$ is the radius of the imaging lens. For the distances $s_{i,o}$ much larger than sizes of both object and image, one can take $\Xi(\vec s, \vec r)\approx 1$.

The representation (\ref{eqn:Gn_thermal}) can be derived from the corresponding property of Gaussian field correlations
\begin{equation}
\begin{gathered}
\left\langle\left[\prod\limits_{i=1}^n E_o(\vec s_{l_i})\right]^{\dagger}\left[\prod\limits_{j=1}^n  E_o(\vec s_{m_j})\right]\right\rangle \\ {} = \sum_{(q_1,\ldots,q_n) \in S(m_1,\ldots,m_n)}\prod\limits_{i=1}^n \left\langle E_o^\dagger(\vec s_{l_i}) E_o(\vec s_{q_i}) \right\rangle.
\end{gathered}
\label{eqn:field_correlation}
\end{equation}

The pairwise correlations, introduced by Eq.~(\ref{eqn:I_general}), can be represented as
\begin{equation}
\label{eqn:I_decomposed_simplified}
I_{ij} \equiv I(\vec r_i, \vec r_j) = \sum_{  l,m} D^{(ij)}(l,m) x_l^\ast x_m,
\end{equation}
where
\begin{equation}
\label{eqn:D_decomposed_simplified}
\begin{gathered}
D^{(ij)}(l,m) = \int_{OP} d^2 \vec s d^2 \vec s' \exp\left(-|\vec s - \vec s'|^2 / w_c^2\right) \\ {} \times d_l(\vec s') [d_m(\vec s)]^\ast h^\ast(\vec s, \vec r) h(\vec s', \vec r').
\end{gathered}
\end{equation}

The coefficients for calculation of $n$th order correlation function at the $k$th set of points ($\vec r_{j_1}$, \ldots, $\vec r_{j_n}$) can be expressed as
\begin{equation}
\label{eqn:coeff_for_Gn_thermal}
\begin{gathered}
D^{(k)}(l_1\ldots l_n;m_1\ldots  m_n) \\ {} = \sum_{(q_1,\ldots, q_n) \in S(1,\ldots,n)} \prod_i D^{(j_i, j_{q_i})}(l_i,m_{q_i}).
\end{gathered}
\end{equation}

Features of the coefficients $D^{(jk)}(l_1,l_2;m_1,m_2)$ can be illustrated with the example of small pixels, placed at points $\vec s_j$ as the functions $d_{j}(\vec{s})$.
For such basis functions, one has
\begin{eqnarray}
\nonumber
D^{(jk)}(m,n)\approx \sigma^2 \exp\left(-|\vec s_m - \vec s_n|^2 / w_c^2\right)\times \\
h(\vec s_m, \vec r_j)h(\vec s_n, \vec r_k),
\label{dtwopoint}
\end{eqnarray}
where $\sigma$ is the  area of the pixel. The PSF $h(\vec s, \vec r)$ diminishes with argument, i.e.~tends to zero for $|\vec s + \vec r s_o / s_i| \ll \Delta l$, where $\Delta l = s_o \lambda/(2 R)$ is the characteristic width of the PSF, determining the Rayleigh limit. The coefficients $D^{(jk)}(m,n)$ will tend to zero with increasing of $|\vec s_m + \vec r_j s_o / s_i|\rightarrow\infty$ or $|\vec s_n + \vec r_k s_o / s_i|\rightarrow\infty$. Our measurement scheme is therefore indeed parametrically local. An especially simple form of those locality restrictions can be obtained for a choice of object pixels in accordance with the used detection positions as $\vec s_j = -\vec r_j s_0 / s_i$: the coefficients $D^{(jk)}(m,n)$ are effectively zero for $|j - m| \ll n_0$ or $|k - n| \ll n_0$, where $n_0 = \Delta l / d$ is the width of the PSF expressed in terms of the object pixel size $d$.

\begin{figure}[htbp]
\centering\includegraphics[width=8.5cm]{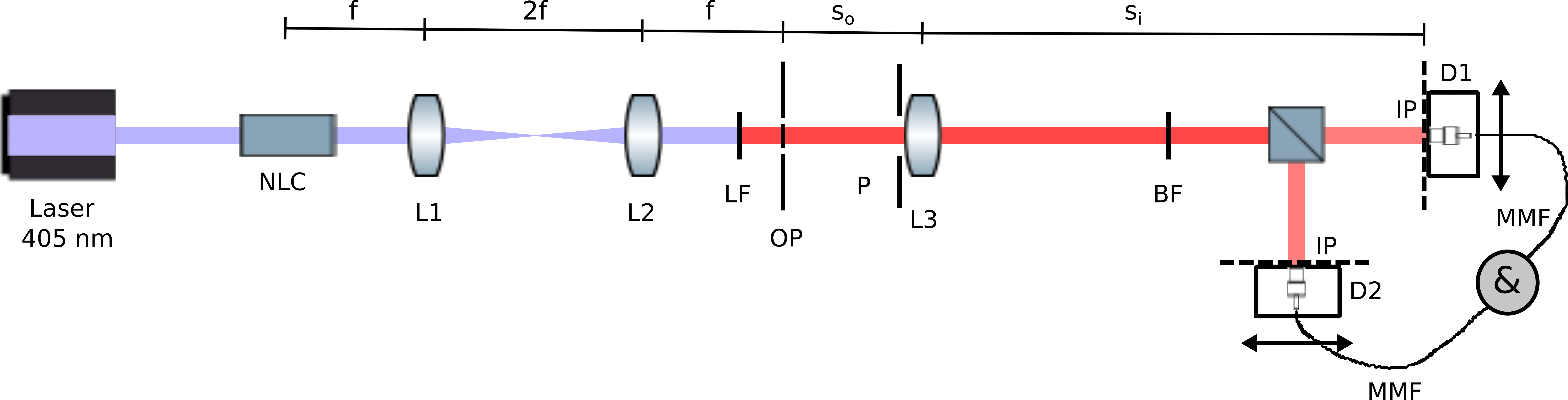}
\caption{SPDC setup. A monochromatic laser is weakly focused into a PPKTP nonlinear crystal (NLC) to generate type-0 SPDC. The two-photon state is imaged via a 4-$f$ arrangement from the center of the NLC to the object plane (OP) using lenses L1 and L2. A long-pass filter (LF) blocks the pump and a band-pass filter (BF) transmits photons at 810\,nm. A single lens imaging system with lens L2 maps the OP onto the image plane (IP) which coincides with the fiber tip of two multimode fibers (MMFs) connecting the detection stages D1 and D2 in a coincidence circuit. The resolution of the imaging system is modified by a variable size pinhole (PH).  \label{fig:setup_SPDC}}
\end{figure}

\subsection*{SPDC entangled photon states}

The setup for imaging with entangled photons is shown in Fig.~\ref{fig:setup_SPDC}.  Our source generates type-0 position-momentum entangled photon states by pumping a 12\,mm long PPKTP nonlinear crystal (NLC) with a continuous wave (CW) laser centered at 405\,nm \cite{walborn10}. The entangled photons are then emitted at 810\,nm. The residual pump beam is subsequently blocked by a long-pass filter (LF) and the subsequent band-pass filter (BF) transmits photons at 810\,nm with a spectral FWHM of 10\,nm to the detectors. The experimental setup contains two imaging systems: The first system consists of a 4-$f$ image using lenses $L1$ and $L2$ both with focal length $f=50$\,mm. This configuration maps the entangled photon states transverse momentum distribution from the OP1 at the center of the NLC to the OP with a magnification factor of $m=1$. The OP is then imaged with a single lens system onto the fiber tips of two multimode fibers (MMFs). Thereby, we have a magnification factor of $m=12$ for $s_o=65$\,mm and $s_i=780$\,mm. Both detection stages can be scanned in horizontal direction.  This setup was used to record the image of a three-slit pattern of a positive USAF resolution chart (Group 4, Element 1) by measuring a second-order correlation function of the photons. The experimental correlation map and the reconstructed object can be seen in Fig.~3
(c),(d) of the main text.

For this case, the probability of having simultaneous clicks of the detectors at the position $\vec{r}_j$ and $\vec{r}_k$ is given by
\begin{eqnarray}
{p}_{jk}\propto |\Phi(\vec{r}_j,\vec{r}_k)|^2,
\label{p3}
\end{eqnarray}
where  $\Phi(\vec{r}_j,\vec{r}_k)$ can be denoted as the two-photon wavefunction \cite{shih} at the detector plane given by
\begin{equation}
\begin{gathered}
\Phi(\vec{r}_j,\vec{r}_k)=\int_{OP}d^2\vec s_1 \int_{OP}d^2\vec s_2 A(\vec s_1)A(\vec s_2)\Lambda(\vec s_1,\vec s_2) \\ {}\times h(\vec s_1, \vec r_j)h(\vec s_2, \vec r_k).
\label{apm1}
\end{gathered}
\end{equation}
  The function $\Lambda(\vec s_1,\vec s_2)$ is denoted as the joint-position amplitude and describes the spatial correlation between photons. Ideally, for the perfectly correlated state $\Lambda(\vec{\rho_1},\vec{\rho_2})\propto \delta(\vec{\rho_1}-\vec{\rho_2})$. In practice, for the used SPDC source, one has  spatial correlations approximately described by 
a finite weight function which also depends on the temperature of the NLC (see, for example, Refs.\cite{walborn10,arie,howell}). Qualitatively, this correlation function can be approximated by a Gaussian function similar as for the pseudo-thermal state in Eq.~(\ref{eqn:I_general}) and the correlation width $w_c$ can be introduced in the same manner.
The dependence in Eq.~(\ref{apm1}) gives us the following expression for the coefficients
\begin{equation}
\nonumber
D^{(jk)}(l_1,l_2;m_1,m_2)=[D^{(jk)}(l_1,l_2)]^*D^{(jk)}(m_1,m_2),
\end{equation}

with

\begin{equation}
\begin{gathered}
D^{(jk)}(m_1,m_2)=\int\limits_{OP}d^2 \vec s_1 \int\limits_{OP}d^2 \vec s_2 d_{m_1}(\vec s_1)d_{m_2}(\vec s_2)\Lambda(\vec s_1,\vec s_2)
\\ {} \times h(\vec s_1, \vec r_j)h(\vec s_2, \vec r_k).
\end{gathered}
\label{dtwo}
\end{equation}
Equation (\ref{dtwo})  has the same structure as Eq.~(\ref{eqn:D_decomposed_simplified}) up to the complex conjugation of some terms and has the same locality properties, the latter imposed by the PSF.

\section*{Supplementary note 4. Optimization of resolution for the twin-photon state}

In the main text,  the optimization of the resolution for our near-field imaging scheme was considered for the pseudo-thermal source.
It was established that there is an optimal correlation width providing for the best resolution. Here we present also results of the simulation for the SPDC twin-photon source and also an existence of the optimal correlation width.
We define the correlation width $w_{c}$ as the FWHM of the spatial correlation function $\Lambda(\vec s) = \Lambda(\vec s, - \vec s)$ described in the Appendix C. Fig. \ref{fig6} shows the dependence of the total variance on $w_{c}$. Similarly to the case of pseudothermal light, there exists an optimal value of the correlation width which has the same order of magnitude as the smallest object features to be resolved. The result is not surprising, if one compares Eqs. (\ref{dtwo}) and (\ref{eqn:D_decomposed_simplified}) which have the same structure and differ by complex conjugation and notations for correlation function only.

\begin{figure}[htb]
\includegraphics[width=\linewidth]{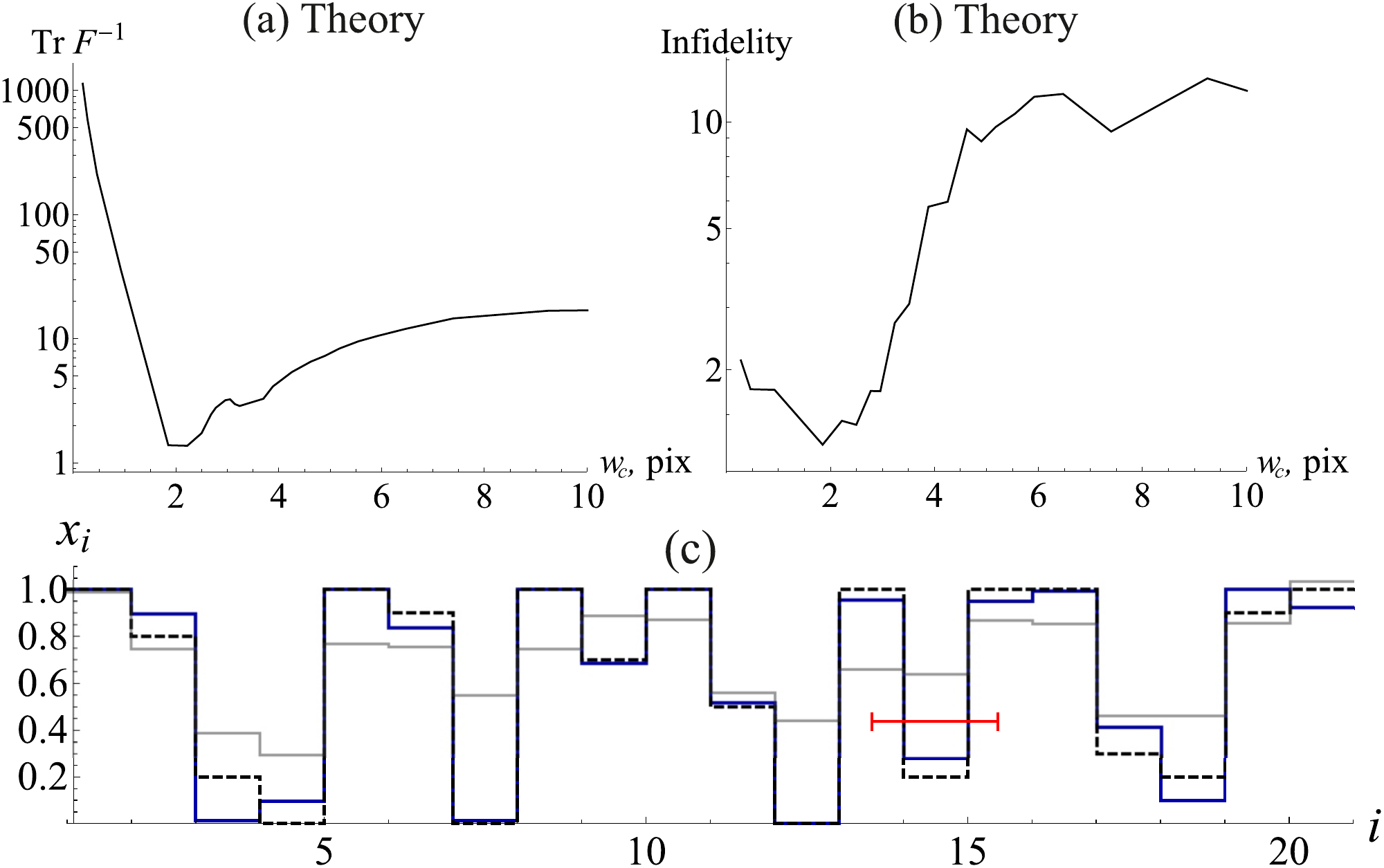}
\caption{Dependence of the inverse Fisher matrix trace (a) and modelled image reconstruction infidelity (b) on the correlation width $w_c$ for the 1D imaging with SPDC source. Result of object reconstruction with optimal correlation width (solid line) is compared to original model object (dashed line) and diagonal part of $G^{(2)}$ (grey line) in the plot (c).
Red  segment corresponds to Rayleigh limit $\Delta l$ in optical system used.}
\label{fig6}
\end{figure}

\section*{Supplementary note 5. Biased estimate}

Here we consider an influence of bias on the bounds for reconstruction errors.  We take the set of $K$ measurements described by the probabilities $p_k(\theta_1,\theta_2,\ldots,\theta_M)$,  for estimation of $M$ real parameters $\theta_m$ subjected to linear inequality  constraints $\theta^{(L)}_m\leq \theta_m\leq\theta^{(H)}_m$. We assume that our reconstruction procedure provides us with the estimator
${\bar \theta}_m(f_1,f_2,\ldots, f_K)$, where $f_k$ are frequencies obtained as the outcome of our measurement device. Our estimator is taken to be biased
\begin{equation}
{\vec b}({\vec\theta};N)=E\left\{{\vec{\bar \theta}}({\vec f})\right\}-{\vec\theta}\neq 0,
\label{bias}
\end{equation}
where $E\left\{\ldots\right\}$ denotes averaging over all realizations of the frequencies, $f_m$, for the fixed total number of measurement runs $N$.
We denote as $\vec x$ the vector $[x_1,x_2,\ldots, x_X]$. Notice that the bias generally depends on $N$.

Using the definition of the bias in Eq.~(\ref{bias}), it is easy to generalize the Cramer-Rao inequality for the lower bound on the covariance  matrix
\begin{equation}
C_{mn}=E\left\{(\theta_m-{\bar \theta}_m)(\theta_n-{\bar \theta}_n)\right\}
\label{cov}
\end{equation}
for biased estimators.
From the Fisher information matrix (\ref{fisher1}), one can generalize the Cramer-Rao inequality deriving the following bound for the elements of the covariance matrix \cite{eldarreview}
\begin{equation}
{\hat C}\succeq \frac{1}{N}({\hat I}+{\hat \Upsilon}){\hat F}^{-1}({\hat I}+{\hat \Upsilon}^T),
\label{cramer1}
\end{equation}
where ${\hat I}$ is an identity matrix, and the ${\hat \Upsilon}$ is the bias gradient matrix with the elements
\begin{equation}
\Upsilon_{jk}=\frac{\partial }{\partial \theta_k}b_j({\vec\theta};N).
\label{grad}
\end{equation}

It is to be noted that the error bound in Eq.~(\ref{cramer1}) depends only on the bias derivative, i.e., the constant  bias does not influence the bound. Also, one can surmise that rapidly varying bias can strongly influence error estimations.
Generally, estimation of the bias is not an easy task. The direct approach would involve estimation of the parameters for a sufficiently large number of realizations of the measurement results obtained for a large number of measurement runs. However, even some rather general considerations about the bias can be used for derivation of the bound. Here we describe the simple case of the bound given in \cite{eldar2004}.

The main concept underlying the derivation of the bound is an existence of a small upper bound for the rate of bias change. Let us assume the possibility to find a number $\gamma<1$ such that
\begin{equation}
\max\limits_{  \phi}{\vec\phi}^T{\hat \Upsilon}^T{\hat \Upsilon}{\vec\phi}\leq\gamma\leq1,
\label{maxgamma}
\end{equation}
where $\phi$ is the normalized vector of $M$  variables, $\phi^T\phi=1$. Then, it is possible to derive the following bound for the total error estimated as the trace of the covariance matrix
\begin{equation}
\mathrm{Tr}\{{\hat C}\}\geq \mathrm{Tr}\{(1-\sqrt{\gamma})^2{\hat F}^{-1}\}.
\label{newbound}
\end{equation}

This simple bound is quite remarkable. It shows that for a non-singular FIM a slowly varying bias is always leading to the improvement of the error (however, notice that it is a biased estimation). Moreover, in the work \cite{eldar2004} it is proven that there is a  penalized maximal likelihood estimation procedure saturating the bound in Eq.~(\ref{newbound}).
Notice that the work \cite{eldar2004} also considers more complicated and refined bounds than Eq~(\ref{newbound}) based on finding the upper bound on some functions of  ${\hat \Upsilon}^T{\hat \Upsilon}$. However, for demonstration of the border resolution enhancement in the realistic imaging scheme the simple bound given by Eq.~(\ref{newbound}) is quite sufficient.

\begin{figure}[htb]
\includegraphics[width=\linewidth]{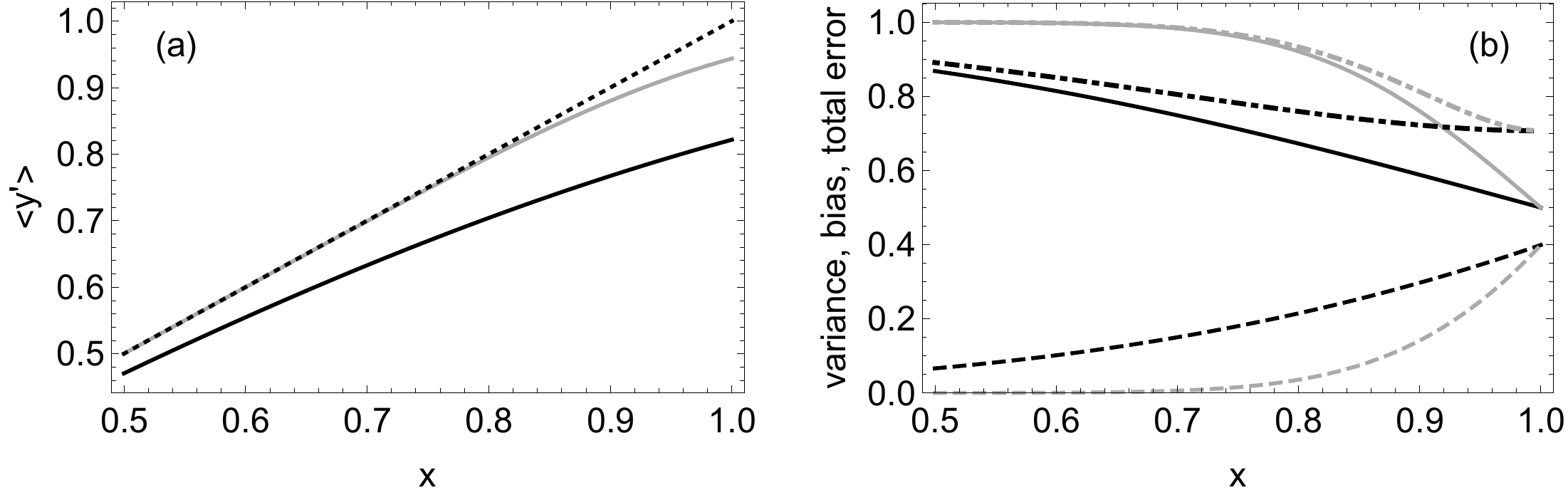}
\caption{Mean value of the biased estimator $y'$ (a). Dotted line shows the unbiased estimator $y$. Variance of the biased estimator $\Delta'$ (solid line), bias $x - \langle y' \rangle$ (dashed line), and the total estimation error $\sqrt{ \langle ( x - y' )^2 \rangle }$ (dot-dashed line) expressed in terms of the variance $\Delta$ of the unbiased estimate are shown in plot (b). Black lines: $F_{11} N = 5$; gray lines: $F_{11} N = 50$.}
\label{fig_biased}
\end{figure}

To illustrate the influence of the estimate bias on its variance, let us consider an example of reconstructing a single parameter $x$ from the measured probabilities. Let $F = (F_{11})$ be the single element FIM for the considered case. For example, for a binary measurement with the probabilities $p(x)$ and $1 - p(x)$ of the two outcomes, the element of the FIM is $F_{11} = \left(\frac{d p(x)}{dx}\right)^2 \frac{1}{p(x)\{1-p(x)\}}$. Let $y$ be the unbiased estimator for $x$, saturating the CRB and characterized by the variance $\Delta^2 = 1 / (F_{11} N)$. For sufficiently large $N$, the values of the estimator $y$ for different realizations of $N$-measurement series are distributed according to the probability density function $w(y) \propto \exp\{-F_{11}N(y-x)^2 / 2\}$.

Now let us consider a constrained problem, where the inequality $x \le 1$ is imposed. To ensure the agreement of the estimated value with that requirement, one can introduce the estimator
\begin{equation}
\label{eqn:constrained_estimator}
y' = \left[ \begin{array}{cc}
              y, & y \le 1, \\
              1, & y > 1.
            \end{array}
 \right.
\end{equation}
The mean value of this estimator equals
\begin{equation}
\begin{gathered}
\left\langle y' \right\rangle = \int_{-\infty}^1 w(y) y dy + \int_1^\infty w(y) dy \\{} = \frac{1}{2}\left(1 - \operatorname{erf} \xi + x \left( 1 + \operatorname{erf} \xi\right) - e^{-\xi^2}\sqrt\frac{2}{\pi F_{11} N}\right),
\end{gathered}
\end{equation}
where $\xi = (1 - x) \sqrt{F_{11} N / 2}$. For $1 - x \gg \Delta$ the mean value tends to the "true" value $x$: $\langle y' \rangle \approx x$, and the estimate is unbiased far from the boundary value 1 (Figure \ref{fig_biased}(a)). For $x \approx 1$ the estimate becomes biased.

The generalized CRB (\ref{cramer1}) for the biased estimator $y'$  gives the value
\begin{equation}
\Delta'^2 \ge \left(\frac{d\langle y' \rangle}{dx}\right)^2 \frac{1}{F_{11} N} = \frac{1}{2}\left(1 + \operatorname{erf} \xi\right) \Delta^2
\end{equation}
for the variance. Far from the boundary, one has $\Delta' \rightarrow \Delta$. However, for $x = 1$ the variance of the biased estimate $y'$ is twice smaller than for the unbiased estimate $y$: $\Delta'^2 = 0.5 \Delta^2$ (Figure \ref{fig_biased}(b)).

The total mean square reconstruction error includes both, the variance and the bias of the estimate, and can be calculated as
\begin{equation}
\langle (y' - x)^2 \rangle = \left[ \frac{1 + \operatorname{erf} \xi}{2} - \xi^2 (1-\operatorname{erf}\xi) - \frac{\xi}{\sqrt{\pi}} e^{-\xi^2} \right] \Delta^2.
\end{equation}
Figure \ref{fig_biased}(b) shows that, regardless of the additional systematic error introduced by the bias, the total reconstruction error for the biased estimator $y'$ is smaller than for the initial unbiased estimator $y$.

The effect can be even more pronounced for multiparameter problems. To illustrate this statement, let us consider a degenerated problem of reconstructing two parameters $x_1$ and $x_2$ when the measured signal depends on their sum $x_+ = x_1 + x_2$ only. In that case, the sum of the parameters $x_+$ can be estimated with a finite error $\delta_+$, $|\delta_+| \le \Delta_+$, while the error of the difference, $x_- = x_1 - x_2$, estimation is unbounded. The errors $\delta_{1,2} = 0.5 (\delta_+ \pm \delta_-)$ of estimating the parameters $x_{1,2}$ themselves remain unbounded as well.

Suppose that now the parameters are bounded from above: $x_{1,2} \le 1$. Therefore, the sum and the difference of the parameters also satisfy the inequalities $x_+ \le 2$ and $|x_-| \le 2 - x_+$. As soon as the "true" values of the parameters reach the corner of the available region, $x_1 = x_2 = 1$, one has $2 - x_+ \le \Delta_+$ and $|x_-| \le \Delta_+$. Now the estimation errors for both, the sum and the difference become finite, and the degenerate problem of finding $x_1$ and $x_2$ becomes solvable with finite accuracy.

It is worth noting, that the predicted position of the optimum at the dependence of the total variance on the correlation remains the same for unbiased (gray objects) and biased (black-and-white objects) estimates.

\end{document}